\documentclass[pdflatex,sn-mathphys-ay]{sn-jnl}

\usepackage{graphicx}%
\usepackage{multirow}%
\usepackage{amsmath,amssymb,amsfonts}%
\usepackage{amsthm}%
\usepackage{mathrsfs}%
\usepackage[title]{appendix}%
\usepackage{xcolor}%
\usepackage{textcomp}%
\usepackage{manyfoot}%
\usepackage{booktabs}%
\usepackage{algorithm}%
\usepackage{algorithmicx}%
\usepackage{algpseudocode}%
\usepackage{listings}%

\theoremstyle{thmstyleone}%
\theoremstyle{thmstyletwo}%

\theoremstyle{thmstylethree}%

\begin{document}

\title[Multiverse analysis, abdication of responsibility and manufacturing of doubt]{Multiverse analysis, abdication of responsibility and manufacturing of doubt}


\author[1]{\fnm{Martin} \sur{Modrák}}\email{modrak.mar@gmail.com}



\affil[1]{\orgdiv{Department of Bioinformatics}, \orgname{Second Faculty of Medicine, Charles University}, \orgaddress{\street{V Úvalu 84}, \city{Prague}, \postcode{15006}, \country{Czech Republic}}}




\abstract{I argue that multiverse analysis is highly suited to two undesirable uses: abdication of researcher's responsibility for their conclusion and manufacturing of doubt. A review of multiverse analyses published in 2025 provides tentative empirical support that abdication of responsibility is present in the literature and I mention anecdotal evidence that multiverse has been used for manufacturing of doubt about Covid-19 precautions. To mitigate negative effects if multiverse analysis becomes widely used I suggest the community adopts two conventions for evaluating multiverse analyzes: evaluating multiverses by the single worst universe they contain and considering large size of a multiverse as a sign of weakness rather than a praiseworthy achievement.}


\maketitle

\section{Introduction}\label{sec:intro}

Multiverse analysis \citep{Steegen-Tuerlinckx-Gelman-etal:2016} is the systematic exploration of results of multiple different specifications of a statistical model, data preprocessing or other steps in an analysis. It is increasingly popular and undeniably useful in some cases --- the author of this text has employed it in his published works to good effect.
I have no doubt that other authors contributing to this special issue will provide nuanced and positive takes on its benefits, drawbacks and applications. I can especially recommend the contribution to this issue by \citeauthor{Rohrer2026-multiverse} (\citeyear{Rohrer2026-multiverse}).  But nuance is not an unalloyed good for understanding the world \citep{Healy2017-nuance} and not everything has to be positive. 

In this paper, I highlight two roles for the multiverse analysis that are clearly socially undesirable, yet multiverse analysis appears very well suited to fill those roles: encouraging researchers to abdicate responsibility for their conclusions and allowing for manufacture of doubt by bad faith actors. I also specifically argue against large multiverses and provide some recommendations for evaluation of claims based on multiverse analyses.

In the rest of this paper, I will employ the same nomenclature as 
used by \citeauthor{Rohrer2026-multiverse} and generally work within the framework they propose. In particular, they distinguish three types of multiverse applications: as an inferential tool, as a persuasion instrument, and as reflection/criticism of a given finding. Further, they discuss the distinction between probabilistic interpretation of the multiverse and a possibilistic interpretation. Under the probabilistic view, we can make inferences from the relative frequencies of certain outcomes in the multiverse (e.g., ``most models show ...'', ``association is positive for over 90\% models``). This is appealing but requires strong assumptions to hold (at least approximately): namely that the individual universes included in the analysis all have  the same probability of being correct and that they are exchangeable\footnote{Or, somewhat more generally that we can formulate a probabilistic model for the distribution over the universes.}. Such strong assumptions are rarely warranted. In contrast, under the possibilistic view, an individual universe only shows that an outcome is \emph{possible} without giving any specific weight to that outcome --- as long as even one universe results in an outcome, the outcome cannot be ruled out. Possibilistic interpretation only requires us to commit to each of the universes being possibly correct, which is a substantially weaker assumption.

Multiverse size is often taken as the headline description of the analysis, typically under the assumption that a larger multiverse is better, either for providing ``more robust'' conclusions or by increasing the probability that at least one of the included universes is correct.

Especially for large multiverses, reporting the results in a suitable way is a challenge of its own. A survey of visualisation types found in the literature is given by \cite{Hall2022-visual-survey}.


\section{Abdication of responsibility}\label{sec:abdication}
The first maladaptive function of multiverse is that it allows --- or even encourages --- the abdication of responsibility of the scientist for the conclusions of their paper. Adding one more option to the multiverse is often easier than evaluating which of the choices are actually sensible. And since larger multiverses are often seen as better, there is an incentive to rather err on the side of including more options than fewer. I do not think this leads to more reliable scientific knowledge. Yes, if there is some fundamental irresolvable ambiguity about a small number analytical choices, checking the sensitivity to such choices is desirable. But if the number of ambiguous choices is large and there is no way to determine which are better, then the scientific conclusion needs to be that the question is not yet ready to be answered and we should instead start laying methodological, theoretical or validation groundwork to determine which choices are good. 
On the other hand, if the choices are in fact not ambiguous and some options are demonstrably better, then the authors should lay out the case for the better choice and not pollute their multiverse with suboptimal universes.
I would only reserve some exceptions to this for the cases where all universes closely agree. 

In a similar vein, a large multiverse can also be used in a sort of ``denial of peer-review attack''\footnote{To my knowledge this term was first coined by Lior Pachter in \url{https://x.com/lpachter/status/1816616175562031120}} --- an attempt to overwhelm a reviewer or any critical reader with the sheer number of analyzes made and results discussed so that they cannot in reasonable time understand them all and thus feel uneasy judging the work negatively.  


The problem is exacerbated when the multiverse is presented in a way that prevents the reader from linking the individual results to their originating universes (e.g., by just summarising frequencies of certain classes of conclusions) and thus the reader is unable to substitute their judgment for the authors.



I should also note that no bad intent on the part of the researcher is required for this problem to arise. As long as multiverse are valued in general and for their size in particular, the researcher may abdicate their responsibility while sincerely believing they are following a good practice.

\section{Manufacturing doubt}\label{sec:ambiguity}

The second maladaptive function of the multiverse is its potential for manufacturing doubt. Whenever an actor wants to increase doubt about a conclusion (e.g., vaccine efficacy to prevent death) they may ran a large multiverse and --- among other analytical options --- allow for inclusion of a single well chosen collider or mediator as a predictor (e.g., hospitalization) and they may easily obtain a multiverse where around half of the effects are close to zero or even of the opposite sign.

The fact that including even one bad binary choice can lead to 50\% of the universes to be wrong has been previously noted \citep{Del_Giudice2021-multiverse-guide}. When there are multiple bad choices, the problem is exacerbated quickly. A particularly striking example is discussed in \cite{Auspurg2025-few-models-better}. In their evaluation of a published multiverse analysis they conclude that ``justified models (n = 1,152) are vastly outnumbered and effectively buried within the unjustified set (n = 91,008)''. 

In addition, all the considerations from the previous section apply. In particular, by using a multiverse, a bad actor needs not commit to a single model and thus can raise the perceived bar for criticism of their position. Including bad or wrong model is not a problem \emph{for them} as they are just asking questions and isn’t that what science is truly about?

\section{Is this an actual problem?}\label{sec:empirical}
I provide some brief evidence to support that the problems identified in this paper arise in the published literature using multiverse analysis. I do not aim for a thorough evaluation and quantification of the prevalence of the problem, instead I am content to show that the problems exist in the literature.

\subsection{Methods}
On 15 February 2026, I searched OpenAlex for “multiverse analysis” (with quotes) and further filtered the results to only works published in 2025. All of the results were checked for duplicates and whether they perform a multiverse analysis of their own (reports on previously published analyzes were excluded as well as multiverse analysis using simulated data only). Descriptive variables as well as several proxies presumed for abdication of responsibility were annotated for each study.

For all results where I could obtain the full text the multiverse was classified into reflective or inferential/persuasive (following \cite{Rohrer2026-multiverse}). The persuasion purpose of multiverse was bundled with inferential, as the distinction between the two lies partly in the intentions of the authors and is not clearly visible in the text. Of note is that any study which reported a primary analysis and then ran a multiverse to assess its robustness was annotated as reflective. When multiple multiverse analyzes were reported in the same paper, the first was chosen. We noted the size of the multiverse and whether the authors claimed that all the results were qualitatively identical and if not, whether authors discuss the main drivers of the observed heterogeneity and make any attempt at evaluating which universe are better or worse (e.g. due to mismatch between model assumptions and observed data), whether they interpret the multiverse probabilistically (e.g. by taking means, reporting percentage of universes meeting some criteria) or possibilistically (could be both), whether one can in principle link individual results to individual universes and whether there is at least one visualization of the multiverse that does not allow the reader to trace the results to individual universes and whether this in fact prevents the reader from learning what is the difference between universes that provide qualitatively different conclusions.

For transparency, I note that the criteria for classifying studies have evolved as I have read through the first quarter of the results and better understood the range of options seen in the literature. The search and the specific criteria were not preregistered.

The full dataset with more detailed  coding instructions and the analysis code is available in the supplementary material.

\subsection{Results}

\begin{figure}
    \centering
    \includegraphics[width=0.8\linewidth]{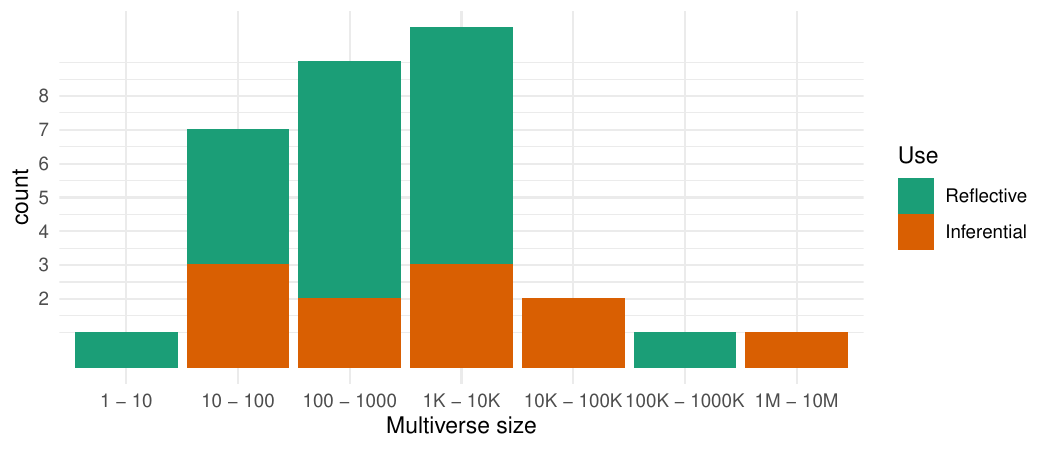}
    \caption{The distribution of multiverse sizes observed in the literature review.}
    \label{fig:size}
\end{figure}

We obtained 83 entries, and removed 17 duplicates, 10 entries where we could not access the fulltext, 14 entries that were not research (e.g. public peer reviews) and 11 that did not perform any previously unreported multiverse analysis. This left us with 31 papers with an original multiverse. In  11 (35.5\%)  studies multiverse was used as an inferential/persuasive tool. 
The size distribution of the multiverses is given in Figure~\ref{fig:size}.
Probabilistic interpretation of the multiverse was very common ---  14 (45.2\%)  studies relied exclusively on a probabilistic interpretation  and further  11 (35.5\%)  studies employed probabilistic interpretation alongside a possibilistic one. In  8 (25.8\%)  studies there was no way to trace reported/visualized results to individual universes and  16 (51.6\%)  studies contained at least one visualization/table where results could not be traced. Of those,  10 (62.5\%)  reported qualitative differences between universes that were indistinguishable in the visualization/table. In 6 (19.4\%)  studies the authors claim that  all universes produced qualitatively equivalent results, of the remaining  22 (88.0\%)  studies highlight which choices they believe are the most important drivers of the  differences in results. Only  7 (22.6\%)  studies make any attempt to evaluate which universes are better than others  (e.g. fit to data, violated assumptions). 

Several of those numbers indicate a possible abdication of responsibility: reporting results that qualitatively differ, but not explaining what drives those differences is a potential problem, made worse by not providing the reader with any way of checking themselves. Another problematic aspect is the reliance on probabilistic interpretation and the lack of checking that some of the models are not mismatched to data. Finally, multiverses with sizes of 1000 or more --- and thus inherently hard to visualize, summarize and understand --- are quite common and were reported in 14 (45.2\%) of the studies. 

We note that the problems we identify are more critical for inferential use of multiverse than for reflective use and so we also present results separately for each of this class in Table~\ref{tab:results}, though we note that the proportions are similar across both use cases.

\begin{table}
\caption{Main results of the literature review on multiverse studies split by the type of multiverse use.\label{tab:results}
}
\begin{tabular*}{\linewidth}{@{\extracolsep{\fill}}lccc}
\toprule
\textbf{Characteristic} & \textbf{All studies} & \textbf{Inferential} & \textbf{Reflective}  \\ 
 & N = 31 & N = 11 & N = 20 \\ 
\midrule\addlinespace[2.5pt]
Probabilistic interpretation only & 14 (45\%) & 7 (64\%) & 7 (35\%) \\ 
Probabilistic and possibilistic interpretation & 11 (35\%) & 1 (9.1\%) & 10 (50\%) \\ 
No way to trace results to universes & 8 (26\%) & 4 (36\%) & 4 (20\%) \\ 
At least one visualization untraceable & 16 (52\%) & 7 (64\%) & 9 (45\%) \\ 
--- of those with qualitative differences & 10 (63\%) & 5 (71\%) & 5 (56\%) \\ 
All results qualitatively identical & 6 (19\%) & 2 (18\%) & 4 (20\%) \\ 
--- of the remaining explain differences & 22 (88\%) & 8 (89\%) & 14 (88\%) \\ 
Post-hoc evaluation of universes & 7 (23\%) & 4 (36\%) & 3 (15\%) \\ 

\bottomrule
\end{tabular*}
\end{table}

A tangential result from this review is that the number of papers found in the search is not very large and multiverse analysis is thus currently not employed very frequently in the literature. In fact many of the multiverses analyzed in this section were primarily used as an example in a methods paper promoting the use multiverse analysis and were not presented as standalone results.
The low number of papers and subjectivity of the classifications is also a reason to not overly generalize the conclusions.

\subsection{Anecdotal evidence for manufacturing doubt}
Whether an analysis aims to manufacture doubt is much harder to evaluate for published papers, but we can provide an anecdote. \citet{goldhaberfiebert2024-covid-critique} critique a published multiverse analysis of the effect of government interventions on the Covid-19 pandemic \citep{Bendavid2024-covid} and note that for many decision points, the multiverse includes choices that are likely invalid. The most problematic in my view is that the tested lags between policy \emph{announcement} (policy implementation date is never used) and reduction in cases/mortality are quite short --- 2 or 4 weeks. Even assuming implementation tracks announcement very quickly (which was not alwasy the case), 4 weeks is way too short for effects on mortality and at best on the shorter end of reasonable lags for effects on reported cases, i.e., at least three quarters of the universes considered are biased to find null results. The multiverse analysis was used to advance the idea that ``we find no patterns in the overall set of models that suggests a clear relationship between COVID-19 government responses and outcomes.'' and the authors specifically argue, that multiverse analysis lets them  ``limit the role of data and model choices in driving the results, or ‘researcher degree of freedom’ ''. They also take a probabilistic interpretation of the multiverse.  Eran Bendavid, the first author on the flawed multiverse has publicly downplayed the pandemic and has been a long-term critic of government interventions against Covid, cooperating among others with Jay Bhattacharya. It is obviously impossible to infer the author's intentions with any certainty, but if \citeauthor{Bendavid2024-covid} wanted to use a multiverse to sow doubt about government interventions while keeping the appearances of best-practice science, they could not do much better.

\section{Discussion}\label{sec:discussion}

I have argued that multiverse analysis provide opportunity for the scientists to abdicate responsibility and for bad actors to manufacture doubt. The data collected suggest that the problem is not purely theoretical, although the full extent is hard to measure precisely.

The two negative roles I highlighted may contribute to multiverse turning out less beneficial for science when more widely adopted than the benefits we see from its early uses, which are likely to predominantly come from careful researchers interested in methods improvement. Though one can argue that this is to be expected for any methodological improvement  \citet{Munafo2026-reform-unwinnable}. And there is hardly any method that cannot be misused.

At the same time, it is natural to try to set social norms and expectations around multiverse analysis to make the undesirable roles harder to present as good scholarship while not diminishing the usefulness of multiverse when carefully employed. I would recommend that when multiverse is used for inferential or persuasive goals, we adopt the methodological convention that, unless proven otherwise, a critique of any single analysis included in the universe should be taken as directly affecting the interpretation of the whole multiverse. In other words, the quality of a multiverse should generally be judged as no higher than the quality of the \emph{worst} universe included. This moves the burden of proof towards the author and --- unless all universes agree --- requires them to fully accept responsibility for all of the universes included. It also makes it easier, not harder to make effective criticism of a multiverse, should they be used in bad faith. 

Additionally, I think it is warranted to see large multiverses as a red flag in their own right: either the scientific question at hand is not mature enough to let us attempt a serious answer or the authors have abdicated their scientific responsibility of actually determining which combinations of modeling, preprocessing or other analysis choices are in fact reasonable. Large size also makes it more likely that a problematic model was included and makes it harder to check quality of all universes.

It is obvious that both recommendations are not absolute and there are cases of useful multiverses that are large and/or contain some problematic analyzes. My argument is that it would be useful for our overall information economy to move more of the burden of proof on authors of multiverse analyses than on the readers trying to understand the multiverse. We should also discourage probabilistic interpretation of multiverses as that makes all of the problems I discuss worse.


My arguments and empirical data are less relevant to the use of multiverse as a reflective tool. If, for example, a multiverse is built from different analytical choices a given group of authors has used to analyze the same type of data in their previous work, then showing the results vary substantially is useful to contextualize a given result, even if some of the analytical choices included are clearly bad. What I think remains relevant is that even for reflective multiverses probabilistic interpretation and inability to track results down to their specific universes are common in the literature, despite being undesirable.




\backmatter

\bmhead{Supplementary information}

A supplementary file containing the annotated studies used in Section~\ref{sec:empirical} as well code to compute the summaries and figure presented.

\bmhead{Acknowledgements}

I thank Jitka Řežábková for help with annotating a subset of the papers for Section~\ref{sec:empirical}. All annotations were finally checked be me and any responsibility for errors is mine.





\bibliography{bibliography}

\end{document}